\documentclass[11pt]{article}
\usepackage{amsmath}
\usepackage{amsfonts}
\usepackage{amssymb}
\usepackage{latexsym}
\usepackage[dvips]{epsfig}
\usepackage{graphicx}
\usepackage{tikz}
\usepackage{etex}
\usetikzlibrary{arrows,backgrounds,snakes}
\usetikzlibrary{shapes.geometric}
\usetikzlibrary{decorations.markings}
\usepackage{lipsum}
\usepackage{multirow}
\usepackage{psfrag}
\usepackage{pstricks}
\usepackage{colordvi}
\usepackage{subcaption}
\usepackage{wrapfig}
\usepackage{amsmath,amsfonts,amssymb}
\usepackage{mathrsfs}
\usepackage{url}
\usepackage{array}

\makeatletter\@addtoreset{equation}{section}\makeatother

\newcommand{\bb}{\mbox{${\bf b}$}}

\newcommand{\bI}{\mbox{\boldmath{$I$}}}

\newcommand{\bn}{\mbox{${\bf n}$}}

\newcommand{\bq}{\mbox{${\bf q}$}}

\newcommand{{\bs}}{\mbox{${\bf s}$}}
\newcommand{\bt}{\mbox{\boldmath{$t$}}}

\newcommand{{\bS}}{\mbox{${\bf S}$}}
\newcommand{\bu}{\mbox{${\bf u}$}}

\newcommand{\bv}{\mbox{${\bf v}$}}

\newcommand{\bx}{\mbox{${\bf x}$}}

\newcommand{\bepsilon}{\mbox{\boldmath{$\varepsilon$}}}

\newcommand{\bsigma}{\mbox{\boldmath{$\sigma$}}}

\newcommand{\bpi}{\mbox{\boldmath{$\pi$}}}
\newcommand{\bPi}{\mbox{\boldmath{$\Pi$}}}

\newcommand{\bzeta}{\mbox{\boldmath{$\zeta$}}}

\newcommand{\bzero}{\mbox{$\bf 0$}}

\newcommand{\BBC}{\mbox{$\mathbb{C}$}}

\newcommand{\BBK}{\mbox{$\mathbb{K}$}}

\newfont{\twelvemsb}{msbm10 at 11.6pt}

\newcommand{\beqn}{\begin {equation}}

\begin{document}
\begin{center}
{\Large {\sc Numerical Investigations of Strain-Gradient Plasticity}}\\
{\Large {\sc with Reference to Non-Homogeneous Deformations}}\footnote{This article was presented at the IUTAM Symposium on Size-Effects in Microstructure and Damage Evolution at Technical University of Denmark, 2018}
\vspace{3ex}\\
N Mhlongo$^{\star 1}$ and BD Reddy$^{1}$ 
\let\thefootnote\relax
\footnote{$^\star$ Corresponding author} 
\let\thefootnote\relax
\footnote{$^1$ Department of Mathematics and Applied Mathematics and Centre for Research in Computational and Applied Mechanics, University of Cape Town, 7701 Rondebosch, South Africa.\\ Emails: mhlnot006@myuct.ac.za,\ daya.reddy@uct.ac.za} 
\end{center}
{\small \section*{Abstract} 
In this work, a higher-order irrotational strain gradient plasticity theory is studied in the small strain regime. 
 A detailed numerical study is based on the problem of simple shear of a non-homogeneous block comprising an elastic-plastic material with a stiff elastic inclusion. Combinations of micro-hard and micro-free boundary conditions are used. The strengthening and hardening behaviour is explored in relation to the dissipative and energetic length scales. There is a strong dependence on length scale with the imposition of micro-hard boundary conditions. For micro-free conditions there is marked dependence on dissipative length scale of initial yield, though the differences are small in the post-yield regime. In the case of hardening behaviour, the variation with respect to energetic length scale is negligible. A further phenomenon studied numerically relates to the global nature of the yield function for the dissipative problem; this function is given as the least upper bound of a function of plastic strain increment, and cannot be determined analytically. The accuracy of an upper-bound approximation to the yield function is explored, and found to be reasonably sharp in its prediction of initial yield.
}

\section{Introduction}
Experiments on metallic specimens at the micro scale (approximately $10\mu$m to $100\mu$m) show significant size-dependence which conventional theories of plasticity are not able to capture. These include experiments on torsion \cite{fleck1994strain}, indentation \cite{begley1998mechanics,swadener2002correlation,stelmashenko1993microindentations,ma1995size}, bending \cite{stolken1998microbend}, and thin film applications \cite{xiang2006bauschinger}. For all these cases there may be different explanations; however, there is general agreement about a size-dependence of hardening (the increase in the stress needed to obtain a given plastic strain with the increase of the energetic length scale) and/or strengthening (the increase of the initial yield stress with the increase in magnitude of the dissipative length scale). There is thus a clear motivation for the inclusion of material length scales for constitutive models at the microscale.

Size-dependent effects may be incorporated in conventional plasticity theories by assuming, for example, that the yield stress depends on the plastic strain and its gradient (see for example \cite{aifantis1984microstructural}). Theoretical models that have become widely studied and adopted include those of Anand and Gurtin \cite{gurtin2004gradient,gurtin2005theory} for rate-dependent materials, and Gudmundson \cite{gudmundson2004unified} and Fleck and Hutchinson \cite{fleck1997strain} for both rate-independent and -dependent materials. In these models, gradient effects are accounted for either through their inclusion in the free energy, or in the flow relation. These are referred to, respectively, as energetic and dissipative models, with the associated length scales having a similar nomenclature. The two models can be shown to lead to distinct size-dependent responses, with energetic models accounting for an increase in hardening with increase in length scale, while for dissipative models the corresponding dependence is with respect to strengthening behaviour. 

A distinctive feature of dissipative models pertains to their behaviour under non-proportional loading. This was first explored in \cite{fleck2014strain}, in which it was shown in the context of simple example problems that the response following a change in the boundary conditions for plastic strain, when applied in the post-yield range, is initially purely elastic. This phenomenon, known as the elastic gap, has been further studied in \cite{carstensen2017some, martinez2016finite,panteghini2016finite}.   

 In the dissipative strain gradient model, the flow relation is given in terms of the microscopic stresses. This cannot be used to determine yield locally, since the microscopic stresses are unknown. It has been shown, however \cite{reddy2011role}, that the flow relation can be expressed in terms of the Cauchy stress through a global formulation using the dissipation function. The form of the flow relation in terms of the yield function and a normality law can then be obtained from a dualization procedure. However, it is not possible to invert this relation in closed form to obtain the generalized plastic strain rate as the normal to a global yield function \cite{carstensen2017some}. Upper bounds to the yield function are explored in \cite{carstensen2017some}, as well as in the recent work \cite{mcbride2018dissipation}, in the context of simple shear. 

The aim of this study is to explore numerically a strain-gradient plasticity model, under conditions of non-homogeneous deformation. The intention is to augment various studies based on problems involving one-dimensional deformation
by examining the implications of variation in deformation, stress, and other variables, in two dimensions. The model used is a rate-independent formulation 
presented in \cite{reddy2008well,reddy2011role}, which is in turn based on the thermodynamically consistent strain-gradient theory of Gurtin and Anand (\cite{gurtin2005theory}). The defect energy is based on Nye's tensor as proposed and adopted in \cite{panteghini2016finite}. 

We investigate  strengthening and hardening; the elastic gap in the case of non-proportional loading; and the global flow relation for the purely dissipative problem. 

The rest of this work is organised as follows. We introduce the governing equations and corresponding weak formulation  in Section \ref{sec:gov}. The formulation for the global flow relation is discussed further in Section \ref{Global-Flow-Relation}. Section \ref{Results} is devoted to the numerical study, and some concluding remarks are presented in Section \ref{Conclusion}.

\section{Governing equations}\label{sec:gov}
We consider a body occupying a domain $\Omega$ with boundary $\Gamma$. 
Assuming quasistatic behaviour, the equation of macroscopic equilibrium is given by 
\begin{equation}
- \mbox{div}\,\bsigma = \bb 
\label{equm}
\end{equation}
in $\Omega$, where $\bsigma$ is the stress and $\bb$ the body force. The boundary conditions are
 \begin{align}
 \pmb{u} = \overline{\pmb{u}} \quad \text{on}~ \partial\Omega_{D},\\
 \pmb{\sigma~ n} = \overline{\pmb{t}} \quad \text{on}~ \partial\Omega_{N},
 \label{Equilibrium BC}
 \end{align}
in which $\bu$ is the displacement, $\bn$ the outward unit normal to $\Gamma$, and 
$\overline{\bu}$ and $\overline{\bt}$ are respectively  a prescribed displacement and traction on $\partial\Omega_{D}$ and $\partial\Omega_{N}$, with $\Gamma= \partial\Omega_{D} \cup \partial\Omega_{N}$ and $\partial\Omega_{D} \cap \partial\Omega_{N} = \emptyset$. 
 
The strain tensor $\pmb{\varepsilon}$ is decomposed into elastic and plastic constituents $\pmb{\varepsilon}^e$ and $\pmb{\varepsilon}^p$, respectively:
 \begin{align}
 \pmb{\varepsilon}=\pmb{\varepsilon}^e + \pmb{\varepsilon}^p.
\end{align} 
 We assume no volume change accompanying plastic behaviour, so that
 \begin{align}
 \text{tr}~\pmb\varepsilon^p = \varepsilon_{ii}^p = 0.
 \end{align}
The elastic relation is
\begin{equation}
\pmb{\sigma} = \BBC(\pmb{\varepsilon} - \pmb{\varepsilon}^p);
\label{elast}
\end{equation}
for isotropic linear elasticity the elasticity tensor $\BBC$ is given by 
\begin{equation}
\BBC\bepsilon = \lambda (\mbox{tr}\,\bepsilon)\bI + 2\mu \bepsilon.
\label{elast}
\end{equation}
Here $\lambda$ and $\mu$ are the Lam\'e parameters, and $\bI$ is the identity tensor.

We define a symmetric and deviatoric microstress $\pmb{\rho}$ power conjugate to the plastic strain rate $\dot{\pmb{\varepsilon}}^p$ and a third-order microstress $\mathbb{K}$ power conjugate to the gradient of plastic strain rate $\nabla \dot{\pmb{\varepsilon}}^p$ \cite{gurtin2005theory}. The quantity $\mathbb{K}$ is symmetric and deviatoric in its first two indices. We also define a defect stress $\pmb{\zeta}$ conjugate to the dislocation density tensor  \cite{gurtin2004gradient}
\begin{align}
\pmb{\alpha} = \text{curl}~\pmb{\varepsilon}^p = \varepsilon_{ikl}~ {\varepsilon}^p_{jl,k}~ \textbf{e}_i \otimes \textbf{e}_j. 
\end{align} 
The generalized stress ${\sf S}$ and plastic strain ${\sf \Gamma}$ are the ordered pairs
\begin{equation}
{\sf S} = (\bpi,L^{-1}\bPi),\qquad
{\sf \Gamma} = (\pmb{\varepsilon}^p, L \nabla \pmb{\varepsilon}^p)\,
\label{SandGamma}
\end{equation}
with magnitudes
\begin{align}\label{Gen. magnitude}
|\mathsf{S}| &= \sqrt{|\pmb{\rho}|^2 + \mbox{$\frac{1}{L^{2}}$} |\mathbb{K}|^2},\\[8pt]\label{Gen. Stress2}
|\mathsf{\Gamma}| &= \sqrt{\textstyle{\frac{2}{3}}|\pmb{\varepsilon}^p|^2 +  \textstyle{\frac{2}{3}}L^2 |\nabla \pmb{\varepsilon}^p|^2}.
\end{align}
Here $L$ is a dissipative material length scale, and the inner product of the two generalized quantities is denoted by
\[
{\sf S}\circ {\sf \Gamma}
  := \pmb{\rho}:\pmb{\varepsilon}^p + \mathbb{K}\circ\nabla\pmb{\varepsilon}^p = \rho_{ij}\varepsilon^p_{ij} 
  + \mathbb{K}_{ijk}\varepsilon^p_{ij,k}\,.
\]
The microscopic stresses and the Cauchy stress  are related to each other through the microscopic force balance equation
\begin{align}\label{microbalance}
\pmb{\sigma}^{dev} - \pmb{\rho} + \text{div}~\mathbb{K} - \text{sym}[\text{dev}(\text{curl} ~\pmb{\zeta})] = \textbf{0} 
\end{align}

We impose higher-order boundary conditions, that is, micro-hard and micro-free boundary conditions on complementary parts $\partial \Omega_H$ and $\partial \Omega_F$ of the boundary, in the form
\begin{align}
\mathbb{K} \pmb{n} + \text{sym}[\text{dev}(\pmb{\zeta} \times \pmb{n})] = \pmb{0} \quad \text{on}~ \partial\Omega_{F},\\
\pmb\varepsilon^p = \pmb{0} \quad \text{on}~ \partial\Omega_{H}.\label{Plastic flow BC}
\end{align}

 The free energy comprises an elastic term $\Phi^e$, a defect term $\Phi^d$ and an isotropic hardening term $\Phi^h$:  
\begin{align}\label{free energy 2} 
 \Phi(\pmb{\varepsilon}, \pmb{\varepsilon}^p, \pmb{\alpha}) &= \Phi^e(\pmb{\varepsilon} - \pmb{\varepsilon}^p) + \Phi^d(\pmb{\alpha}) + \Phi^h(\eta).
\end{align}
Here $\Phi^d(\pmb{\alpha})$ is the defect energy 
\begin{align}
\Phi^d(\pmb{\alpha}) = \mu l^2~ \pmb{\alpha}\cdot \pmb{\alpha} ,
\label{Phid}
\end{align}
where $l$ is an energetic material length scale, and $\eta$ is a hardening parameter to be specified. 
The free-energy imbalance takes the form
 \begin{align}\label{Local dissipation inequality}
 \dot{\Phi} - \pmb{\sigma}: \dot{\pmb{\varepsilon}}^e - \pmb{\rho}:{\dot{\pmb{\varepsilon}}^p} - {\mathbb{K}}:\nabla \dot{\pmb{\varepsilon}}^p -  \pmb{\zeta}: {\dot{\pmb{\alpha}}} \leq 0.
 \end{align}
 Use of the elastic relation \eqref{elast}, and the definitions
\begin{align}
\pmb{\zeta} = \dfrac{\partial \Phi^d}{\partial \pmb{\alpha}}
\label{zeta}
\end{align}
 and 
\begin{align}\label{g}
  g = -\dfrac{\partial \Phi^h}{\partial \eta},
\end{align}   
 leads to the reduced dissipation inequality
%
%
%
\begin{align}\label{reduced dissipation inequality}
\pmb{\rho}:{\dot{\pmb{\varepsilon}}}^p + {\mathbb{K}}:\nabla \dot{\pmb{\varepsilon}}^p +  g~ \dot{\eta} \geq 0,
\end{align}
which forms the basis for construction of an associative flow relation. 


\paragraph{Flow relation}\ \  We define a convex yield function $f$, a function of the dissipative generalized stress $\mathsf{S}$ and the conjugate hardening variable $g$. The set of admissible generalized stresses is then defined to be those values of $\mathsf{S}$ that satisfy the generalized Mises-Hill condition
\begin{align}
\label{yield-function}
f(\mathsf{S,g}) = |\mathsf{S}| + (g-\sigma_0) \leq 0,
\end{align} 
in which $\sigma_0$ is the initial yield stress. 
The flow relation in local form is then 
\begin{subequations}
 \begin{align}
 \dot{\mathsf{{\Gamma}}} &= \lambda \frac{\partial f}{\partial \mathsf{S}} = \lambda \frac{\mathsf{S}}{|\mathsf{S}|},\\
 \dot{\eta} &= \lambda \frac{\partial f}{\partial g} = \lambda\\
 \lambda \geq 0, \quad &f \leq 0 , \quad \lambda f = 0,
 \end{align}
 \label{flow relation}
 \end{subequations}
 where $\lambda$ is a scalar multiplier. 
The flow relation may be expressed alternatively and equivalently in the terms of the convex and positively homogeneous dissipation function $D$, given by 
 \begin{align}\label{Dissipation function SGP}
 \mathcal{D}(\dot{\mathsf{\Gamma}}) = (\sigma_0-{g}) |\dot{\mathsf{\Gamma}}|;
\end{align} 
then we have 
\begin{align}
\mathsf{S} &= \frac{\partial \mathcal{D}}{\partial \dot{\mathsf{\Gamma}}}\label{Obtained 1-1}\\[8pt]
 &=  (\sigma_0 - g)\frac{\dot{\mathsf{\Gamma}}}{|\dot{\mathsf{\Gamma}}|} , \quad \dot{\mathsf{\Gamma}} \neq 0.\label{Obtained 1}
\end{align}


Following \cite{panteghini2016finite}, we set
\begin{align}
\dot{\eta} = |\dot{\mathsf{\Gamma}}|\,,\qquad g = -h\eta^{nh}
\end{align}
in which $h$ is the hardening modulus and $n$ is a non-negative parameter. It follows from (\ref{g}) that 
\begin{align}
\Phi^h = \frac{h}{nh +1} \eta^{nh + 1}.
\end{align}
Later, in the numerical simulations we approximate the dissipation function by \cite{panteghini2016finite}
\begin{align}\label{Chosen Dissipation}
\mathcal{D}_{\delta}(\eta, \dot{\eta}) = \begin{cases}
\displaystyle (\sigma_0 - g) \frac{\dot{\eta}}{2 \delta} \quad \quad &\text{if}~ \eta \leq  \delta\,, \\[6pt]
\displaystyle (\sigma_0 - g) \left( 1 - \frac{\delta}{2\dot{\eta}}\right) \quad \quad &\text{if}~ \eta >  \delta\,,
\end{cases}
\end{align}
 where $\delta$ is a reference strain rate. 
 
 \paragraph{Weak formulations}\label{WeakForm}
 We define spaces of displacements $V$ and of plastic strains $Q$ by
\begin{align*}
 V & =    \{ \bv\ |\ v_i \in H^1(\Omega),\ \bv = \bzero\  \mbox{on}\ \Gamma_D \}\,,
  \\
Q & =   \{ q_{ij} \in H^1(\Omega), \ \ q_{ij} = q_{ji},~ q_{ii} = 0~ \text{and} ~ \textbf{q}= \textbf{0}~\text{on}~ \partial\Omega_{H}\}\,.
\end{align*}
Here $H^1(\Omega)$ is the Sobolev space of functions which together with their first derivatives are square-integrable on $\Omega$.
%
The weak form of the equilibrium equation is standard: find $\bu \in V$ that satisfies
\begin{equation}
\int_\Omega \bsigma :\bepsilon (\bv)\ dx = \int_\Omega \bb\cdot\bv\ dx + \int_{\Gamma_N} \overline{\bt}\cdot\bv \ ds
\label{weakequil}
\end{equation}
 for all $v\in V$. The weak form of the microforce balance equation  is obtained by taking the inner product of \eqref{microbalance} with arbitrary $\bq \in Q$, integrating, and integrating by parts the term involving $\BBK$: this yields
 \begin{align}\label{weak-Microforce-999}
 \int_{\Omega} \big\{\pmb{\sigma}:\textbf{q} - \mathsf{S} \circ \mathsf{Q}\big\} dx - \int_{\Omega}  \text{curl} ~\pmb{\zeta}: \textbf{q}~ dx+ \int_{\partial\Omega} \mathbb{K} \textbf{n} \cdot \textbf{q} ~ds=0.
 \end{align}
 Here we have also used the fact that $\bq$ is symmetric and deviatoric. Integrating by parts the middle term of (\ref{weak-Microforce-999}), we obtain
 \begin{align}\label{Weak-Microforce-9999}
  \int_{\Omega} \big\{\pmb{\sigma}:\textbf{q} - \mathsf{S} \circ \mathsf{Q}\big\} dx + \int_{\Omega}  \pmb{\zeta}: \text{curl} ~\textbf{q}~ dx+ \int_{\partial\Omega}[\mathbb{K} \textbf{n}+ \text{sym}[\text{dev}(\pmb{\zeta}\times\textbf{n})]] \cdot \textbf{q} ~ds=0.
 \end{align}
Using the boundary condition \eqref{Plastic flow BC} and the relations \eqref{Phid} and \eqref{zeta} we obtain the weak form of the microforce balance equation: 
%
 \begin{align}\label{Weak}
 \int_{\Omega} \big\{\pmb{\sigma}:\textbf{q} - \mathsf{S} \circ \mathsf{Q}\big\} dx - \int_{\Omega} \big\{ ~\mu l^2 \text{curl}~ \pmb{\varepsilon}^p:  \text{curl} \textbf{q} \big \}~ dx=0.
 \end{align} 
 Finally, we substitute for $\mathsf{S}$ using the regularized dissipation function in \eqref{Chosen Dissipation} to get 
 \begin{align}\label{Weak-Microforce3}
  \int_{\Omega} \bigg\{\pmb{\sigma}:\textbf{q} -\bigg( \frac{\partial \mathcal{D}_{\varepsilon}}{\partial \dot{\mathsf{\Gamma}}}\bigg) \circ \mathsf{Q}\bigg\} dx - \int_{\Omega} \big\{ ~\mu l^2 \text{curl}~ \pmb{\varepsilon}^p:  \text{curl} \textbf{q} \big \}~ dx=0.
  \end{align}
 The pair of equations \eqref{weakequil} and \eqref{Weak-Microforce3} constitute the weak formulation of the strain gradient problem. 
 
With a view to discrete approximations of the problem we discretize \eqref{Weak-Microforce3} in time. The time interval of interest $[0,T]$ is partitioned with  $0=t_0 < t_1< \cdots < t_N=T$. Time derivatives are replaced by their backward Euler approximations. We denote a time increment at $k+1$ by $\Delta t= t_{k+1} - t_k$. Then from  (\ref{Gen. Stress2}) we have

\begin{equation}\label{Gen. Stress 4}
|\Delta \mathsf{\Gamma}| = \sqrt{\textstyle{\frac{2}{3}}|\Delta \pmb{\varepsilon}^p|^2 +  \textstyle{\frac{2}{3}}L^2 |\Delta \nabla \pmb{\varepsilon}^p|^2},
\end{equation}
so that (\ref{Weak-Microforce3}) becomes
 \begin{align}\label{Weak-Microforce4-time-t}
 \int_{\Omega} \bigg\{{\pmb{\sigma}}^{k+1}:\textbf{q} ~~- \bigg(\frac{\partial \mathcal{D}_{\varepsilon}}{\partial \Delta \mathsf{\Gamma}}\bigg)^{k+1} \circ \mathsf{Q} ~~-\mu l^2 \text{curl}~ \pmb{\varepsilon}^p{}^{({k+1})}: \text{curl} \textbf{q} \bigg \}~ dV =0.
 \end{align}
The stress $\bsigma^{k+1}$ is evaluated using the elastic relation \eqref{elast}.

\section{Global flow relation}\label{Global-Flow-Relation}
The yield condition \eqref{flow relation} is expressed in terms of the indeterminate generalized stress $\mathsf{S}$, so that the yield condition cannot be determined locally from (\ref{Obtained 1}). This is resolved by formulating the flow relation in global form, together with microforce balance, and with the flow relation written in terms of the dissipation function. The result is \eqref{Weak-Microforce3} or, for the time-discrete problem, \eqref{Weak-Microforce4-time-t}.  

The question that then arises is the following: how does one invert this relation to obtain a global flow relation in terms of a yield function? This issue was investigated in \cite{carstensen2017some} for the purely dissipative problem, that is, the problem with $\bzeta = \bzero$ in the present context. The starting point for such an investigation is the global dissipation functional $j(\mathsf{Q})$, defined by 
\begin{equation}
j(\mathsf{Q}) = \int_{\Omega} \mathcal{D} (\mathsf{Q}) dx .
\label{j}
\end{equation}
For convenience we set $\pmb{\zeta}=\textbf{0}$, and define 
\[
\Sigma = (\bsigma, \bzero )\,.
\]
Then \eqref{Weak-Microforce3} can be written in the form
\begin{equation}
\int_\Omega \left( \mathsf{\Sigma} - \frac{\partial D}{\partial \dot{\mathsf{\Gamma}}} \right)\circ \mathsf{Q}\ dx = 0\,.
\label{weakflowD}
\end{equation}
It is important to note that this weak formulation of the flow relation does {\em not} imply the local relation $\mathsf{\Sigma} = \partial D/\partial \dot{\mathsf{\Gamma}}$, which may be inverted to obtain \eqref{flow relation}.

For the global relation we have to follow a different route, and use the property that the global yield and dissipation functions are polar conjugates (see \cite{carstensen2017some}): that is, the global yield function $\Phi (\mathsf{\Sigma})$ may be obtained from
 \begin{align}
 \label{Polar-conjugate}
\Phi(\mathsf{\Sigma}) & =
\text{sup}_{\mathsf{Q \neq 0}}   \frac{\displaystyle\int_\Omega \mathsf{\Sigma} \circ \mathsf{Q}~ dx}{j(\mathsf{Q})} 
\nonumber \\
& = 
\text{sup}_{\mathsf{Q\neq 0}}  \frac{\displaystyle \int_\Omega \pmb{\sigma}:\textbf{q}~ dx}{j(\mathsf{Q})}\,.
\end{align}
The global yield fuinction $\Phi$ is convex and positively homogeneous. Unfortunately, this function cannot be obtained in closed form \cite{carstensen2017some}. In Section \ref{Global-Flow-Relation-Results} we will explore an approximation, in the context of the discrete problem.

\section{Numerical investigation}\label{Results}
In this section we carry out a numerical investigation of the response of a composite rectangular block subject to simple shear, using the strain gradient theory presented earlier. The model problem is designed to have a non-homogeneous response, which allows insights beyond those obtained for homogeneous problems such as an infinite strip in tension or shear (see for example \cite{chiricotto2016dissipative,fleck2014strain,panteghini2016finite}.

The block has height $H=20$mm and width $W=55$mm. The bottom surface is constrained against displacement while the top surface is subjected to a prescribed uniform displacement $u_x=\Gamma H$ in which the applied
shear $\Gamma$ is applied as the increments $\Delta{\Gamma} = 1s^{-1}$.
 In addition, the top surface is constrained against displacement in the $y$-directon ($u_y = 0$). 
The sides of the block are traction-free, and plane strain conditions are assumed. 

Three types of higher-order boundary conditions (BCs) are applied:

\begin{itemize}
\item Microfree BCs - plastic flow is unconstrained on all boundaries; that is, $\mathbb{K}\bn + \text{sym}[\text{dev}(\bzeta \times \bn)] = \bzero$;
\item Microhard BCs - no plastic flow on all boundaries for the duration of the analysis period; that is, $\bepsilon^p = \bzero$;
\item Passivation BCs - Plastic flow is unconstrained on all boundaries for a time interval, after which microhard conditions are applied on all boundaries.
\end{itemize}
Analyses have a duration of $1$ second. The parameters used are as listed in Table \ref{Parameters} below, unless otherwise stated. 

\begin{figure}[!h]
\centering
\includegraphics[scale=0.5]{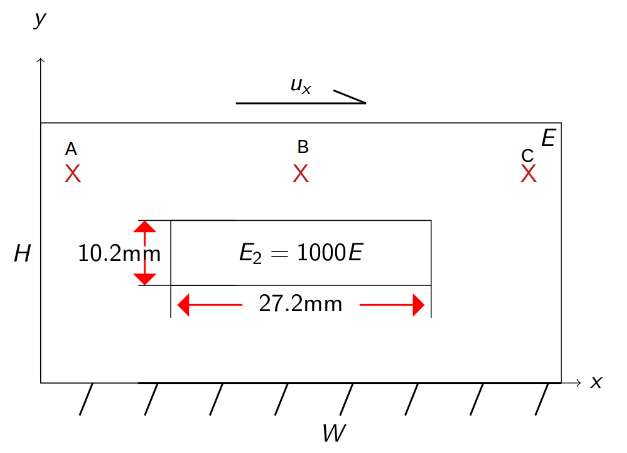}
\caption{The simple shear problem for a non-homogeneous block}\label{Shear-simple-NonH}
\end{figure}

\begin{table}[h!]
\centering
 \begin{tabular}{|l c c c|} 
 \hline
 Parameter ~&& Value & Units \\ [0.5ex]
 \hline 
 Young's modulus $E$ ~&& $68380$ & MPa  \\ 
 Initial yield stress, $\sigma_0$ ~&& $2500$ & MPa\\
 Hardening modulus $H$ ~&& $437.34$ & MPa\\
 reference strain rate $\delta$ ~&& $5\times 10^{-4}$ & s$^{-1}$ \\
 sensitivity parameter $n$~ && $0.2$ &-\\  
 Hardening modulus $h$ &&437.34&MPa \\
 Poisson's ratio $\nu$ && $0.3$ & - \\[1ex]
  \hline
\end{tabular}
 \caption{Material parameters used for the simple shear problem}
 \label{Parameters}
\end{table}
The non-homogeneous body shown in 
Figure \ref{Shear-simple-NonH} has two sections: a purely elastic rectangular inclusion, and surrounding material with the elastoplastic properties listed in Table \ref{Parameters}. The inclusion has Young's modulus $E_2 = 1000 E$. A range of results will be presented using this model problem, though in Section \ref{Elastic gap}, on the elastic gap, for simplicity we will present results for the homogeneous block which has the elastoplastic properties listed in Table \ref{Parameters} throughout the domain. 

Four-noded quadrilateral elements with bilinear approximation of both displacement and plastic strain were used; a mesh comprising $50 \times 50$ elements for the homogeneous block and $51 \times 51$ elements for the non-homogeneous domain was found to provide results of acceptable accuracy.

\subsection{Stress distribution}\label{Stress-Distribution}
To observe the variation in stresses across the domain, we plot the stress components 
and the norm of the deviatoric stress $|\bsigma^{\text{dev}}|$ along the line $ y=0.75H$ (the line connecting sampling points A, B and C in the non-homogeneous block). We consider purely dissipative conditions with $L=0.2H,~ l=0$, and we present results at $0.2$s.
\begin{figure}[!h]
\centering
\includegraphics[width=0.6\textwidth]{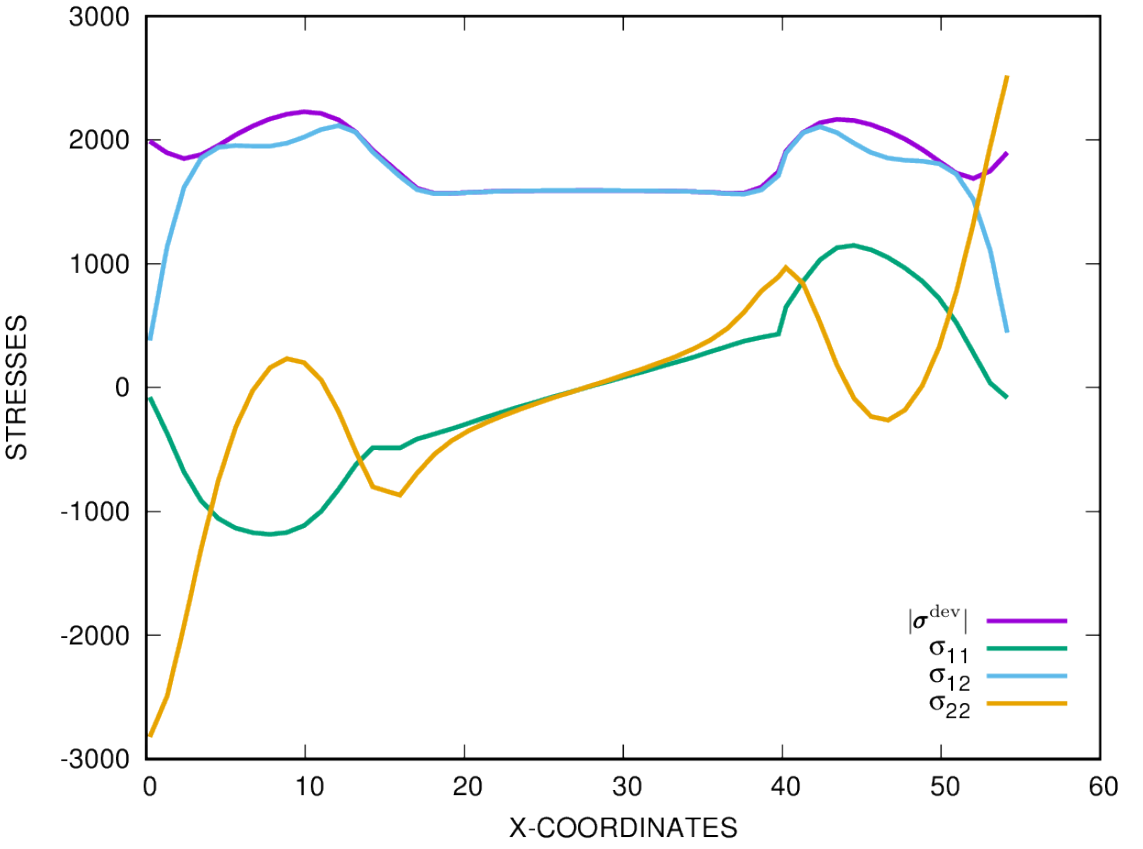}

(a)

\makebox[1cm]{ }\\
\includegraphics[width=0.6\textwidth]{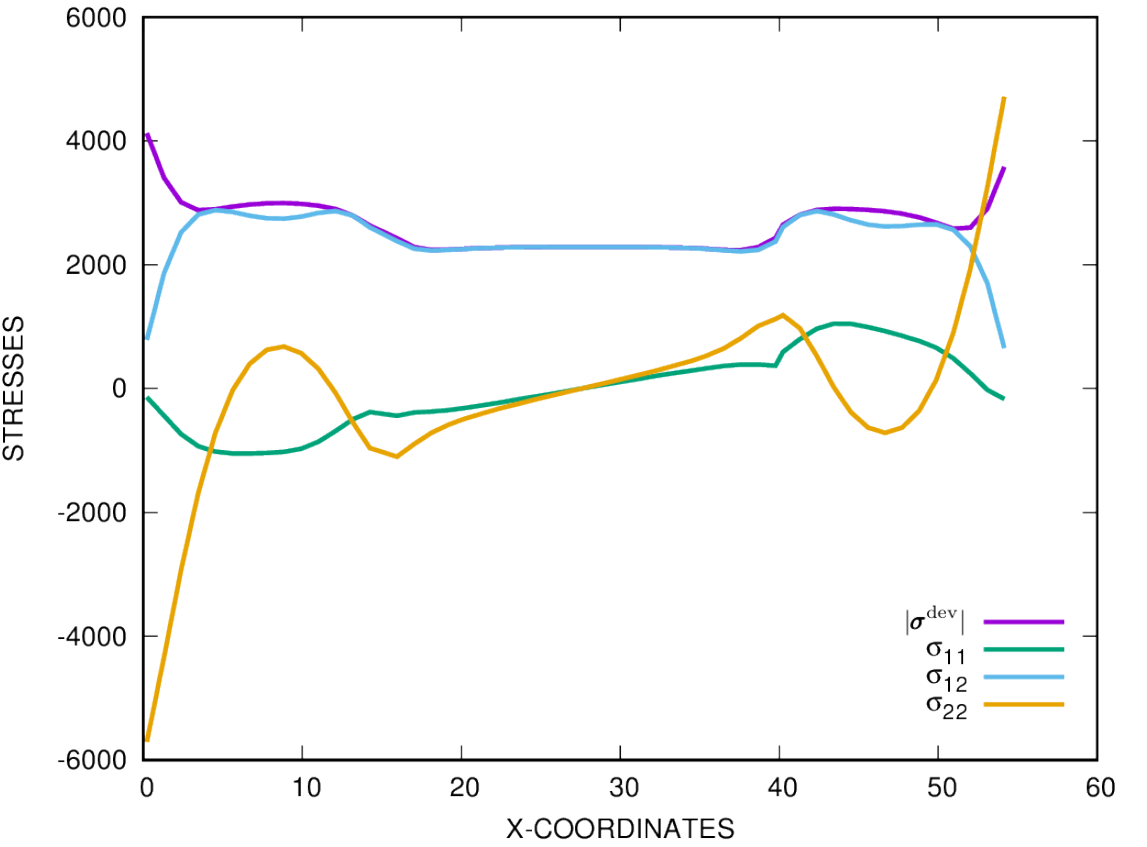}

(b)

\caption{Variation of stresses for the non-homogeneous block along $y=0.75H$ with $L=0.2H, ~l=0$, for (a) microfree and (b) microhard boundary conditions}
\label{Stress-x-variation}
\end{figure}

Figure \ref{Stress-x-variation}(a)
shows the stress variations for microfree BCs whilst Figure \ref{Stress-x-variation}(b) shows results for microhard BCs.
Along the line of symmetry, the direct stresses are zero for both results and the stress $\sigma_{22}$ has maxima and minima at the boundaries. The maximum magnitude of $|\bsigma^{\rm dev}|$ occurs closer to the sides.
All stress magnitudes are significantly higher for the microhard BC results as compared to their corresponding magnitudes in the microfree analysis. This is expected because the microhard BC causes dislocations to pile up on the edges, unlike the microfree BCs which allow dislocations to exit freely.


\subsection{Elastic gap}\label{Elastic gap}
\begin{figure}[!h]
  \centering
    \includegraphics[width=0.6\textwidth]{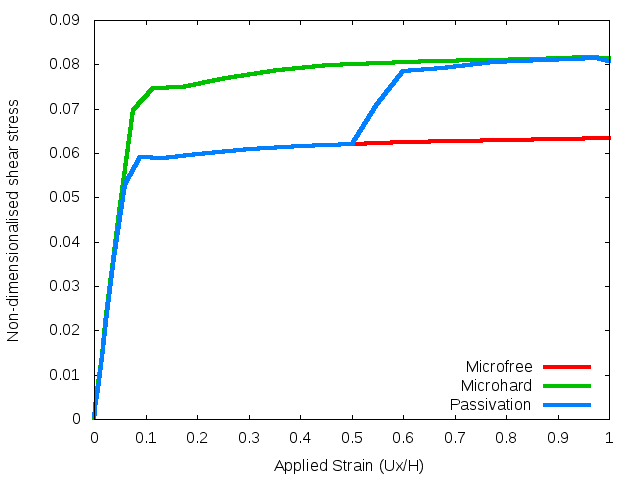}

(a)

    \includegraphics[width=0.6\textwidth]{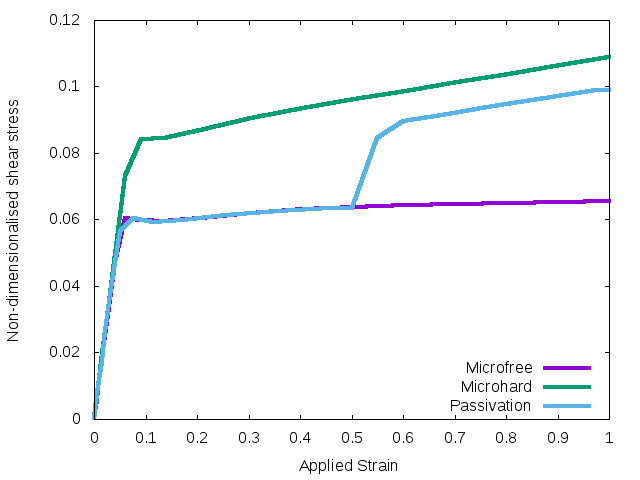}

(b)

  \caption{Passivation, microfree and microhard results using pure dissipative conditions ($L=0.2H$, $l=0$): (a) Homogeneous block; (b) Non-homogeneous block.}
  \label{Meshes-PassivM}
\end{figure}

We illustrate the elastic gap phenomenon, using for this purpose both the homogeneous block, that is, without the elastic inclusion and the non-homogeneous block as defined earlier. Results for the three types of boundary conditions are shown in Figure \ref{Meshes-PassivM}: microfree, microhard,  and microfree for the first $0.5s$ then microhard for the last $0.5s$ are used for this test with the material length scales $L=0.2H$ and $l=0$. Results are extracted from sampling point B. The imposition of zero plastic strain at $0.5s$ in the last boundary condition is referred to as passivation.


For both blocks, the simulations corresponding to microhard and microfree BCs lead to quite distinct responses, as expected, whilst the curve corresponding to passivation shows behaviour in line with the elastic gap phenomenon, though with slopes somewhat smaller than the elastic slope. This departure could be ascribed to the use of a viscoplastic regularization of the rate-independent dissipation function. Moreover, for the non-homogeneous block we observe that even though the passivation curve does not reach the microhard curve within the range considered, the elastic gap has a slope that is much closer to the elastic slope as compared to the homogeneous block.

\subsection{Strengthening and hardening}\label{Diss}
Strengthening refers to the increase  of  the limit of proportionality, or the threshold for the onset of plastic flow, whilst hardening is associated with the increase in the stress required to obtain a given plastic shear. In the context of strain gradient theories, strengthening is generally associated with dissipative models while hardening depends on the magnitude of the energetic length scale (see for example \cite{chiricotto2016dissipative,fleck2015strain,polizzotto2010strain}) for investigations in the context of problems undergoing homogeneous deformation). These features are illustrated here for the model problem, with stress values sampled at point B in Figure  \ref{Shear-simple-NonH}.
 
\begin{figure}[!h]
  \centering
    \includegraphics[width=0.6\textwidth]{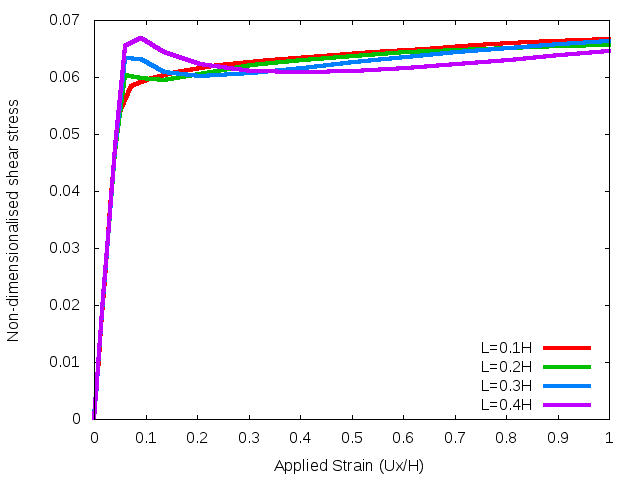}

(a)

    \includegraphics[width=0.6\textwidth]{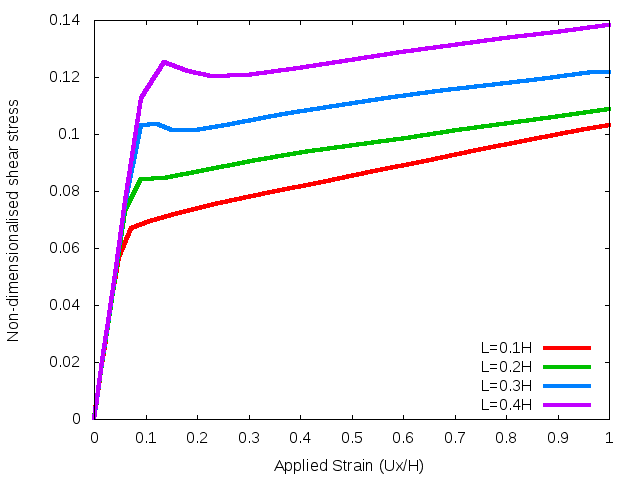}

(b)

  \caption{Shear stress response at point B in Figure \ref{Shear-simple-NonH}, showing strengthening in relation to dissipative length scale: (a)  microfree boundary conditions; (b) microhard boundary conditions}
  \label{Strengthening_mhard_mfree}
\end{figure}

Figures \ref{Strengthening_mhard_mfree} show results of shear stress vs applied strain, for the purely dissipative problem. The microfree results show a significant dependence of the limit of proportionality on length scale; however, beyond the region of initial yield there is little difference in response. In contrast, with microhard boundary conditions there is a strong dependence of initial yield on length scale, with this dependence persisting well into the plastic range.



\begin{figure}[!h]
  \centering
    \includegraphics[width=0.6\textwidth]{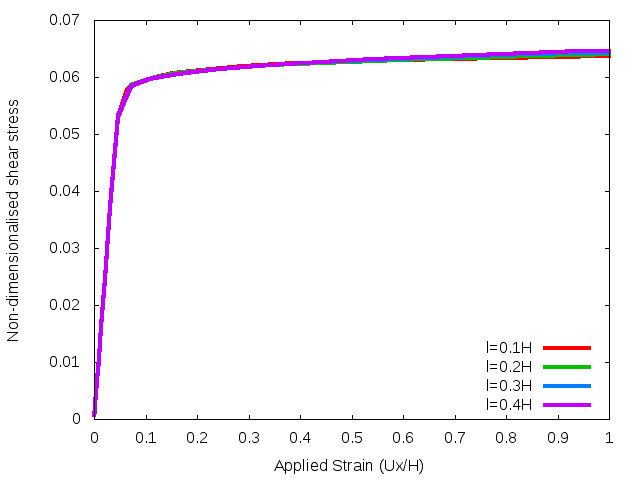}

    (a)
    
    \includegraphics[width=0.6\textwidth]{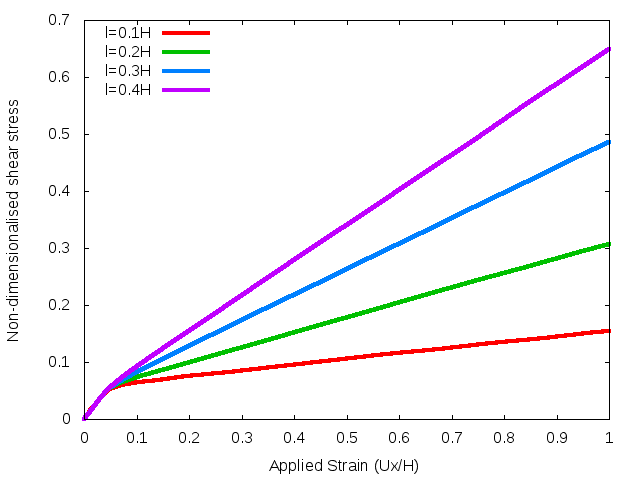}

(b)

\caption{Shear stress response at point B in Figure \ref{Shear-simple-NonH}, showing hardening in relation to the energetic length scale: (a)  microfree boundary conditions; (b) microhard boundary conditions}
  \label{Microfree-Energetic}
\end{figure}

Next, we study the behaviour with respect to variation in energetic length scale magnitudes, with the dissipative length scale set to $L=0$. We present results in Figure \ref{Microfree-Energetic}, for the cases of microfree and microhard BCs.
Results for the microfree boundary condition show insignificant differences in response, for various length scales, while for microhard boundary conditions there is a clear relationship between energetic length scale and the degree of hardening, that is, the slope of the stress-strain curve in the plastic range.

\subsection{Approximation of the global flow relation}\label{Global-Flow-Relation-Results}
In Section \ref{Global-Flow-Relation} we formulated an expression for the global flow relation as a function of the Cauchy stress $\bsigma$. The supremum or least upper bound that characterises the yield function $\Phi$ in equation \eqref{Polar-conjugate} cannot be determined in closed form. 

Turning to a finite element approximation of \eqref{Polar-conjugate}, we make use of the conforming approximations of the displacement $\bu$ and plastic strain $\bepsilon^p$ that form the basis for the results in this section. We set 
\begin{equation}
\bepsilon^p = {\sf Np}\,,\quad \nabla\bepsilon^p = {\sf Bp}\,,\quad \bu = \overline{\sf N}{\sf d},\quad \bepsilon (\bu) = \overline{\sf B}{\sf d},
\label{findim}
\end{equation}
where ${\sf p}$ and ${\sf d}$ are respectively the global degrees of freedom of $\bepsilon^p$ and $\bu$, 
${\sf N}$ and $\overline{\sf N}$ are matrices of shape functions, and ${\sf B}$ and $\overline{\sf B}$ matrices of shape function derivatives. Then \eqref{Polar-conjugate} becomes, for the discrete problem,
\begin{equation}
\Phi ({\sf s}) = \sup_{{\sf q \neq 0}} \frac{{\sf q^Ts}}{{\cal J}({\sf q})}\,.
\label{discrpolar}
\end{equation}
Here the discrete form of the global dissipation function, as a function of arbitrary degrees of freedom ${\sf q}$, is given by
\begin{equation}
{\cal J} ({\sf q})  = \sigma_0  \sqrt{{\sf p}^T{\sf K}{\sf p} },
\label{D2discr}
\end{equation}
where the pointwise matrix ${\sf K}$ is defined by 
$
{\sf K}(\bx) = {\sf N^TN + \ell^2B^TB}\,,
$
and the global vector of nodal stresses ${\sf s}$ is given by 
\begin{equation}
{\sf s} := \int_\Omega {\sf N^T}\mbox{dev}\,\bsigma\,dx\,.
\label{sdef}
\end{equation}
Explicit determination of the least upper bound on the righthand side of \eqref{discrpolar} would give the discrete version of the yield function in terms of the global vector of nodal stresses. Unfortunately, this cannot be evaluated in closed form. It has been shown in \cite{carstensen2017some} that one has the upper bound 
\begin{equation}\label{Upper-Bound}
 \Phi(\mathsf{s}) \leq \sigma_0^{-1}\text{max}_{x\in\Omega}|[{\sf K}(\textbf{x})]^{-1/2}|.
 \end{equation}

Here we  explore numerically an upper-bound approximation to the global yield function for the discrete problem, by choosing ${\sf q} = {\sf s}$ in \eqref{discrpolar}: this gives
 \begin{align}
 \Phi(\mathsf{s}) \leq \overline{\Phi}(\mathsf{s}):= \frac{|\mathsf{s}|^2}{\mathcal{J(\mathsf{s})}}.
 \label{Phibar}
 \end{align}
The function $\overline{\Phi}$ would be expected to predict first yield earlier than when it actually occurs.

We present results on the global yield approximation in Figure \ref{Phibar}, for the case of microhard boundary conditions with purely dissipative behaviour; that is, $L=0.2H$, $l=0$. 
\begin{figure}[!h]
  \centering
 \includegraphics[width=0.6\textwidth]{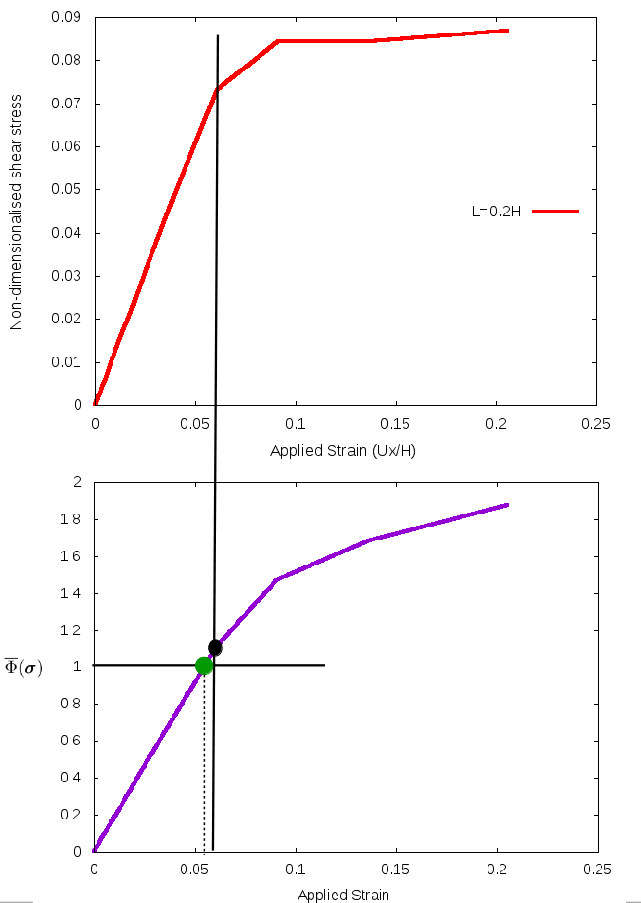}
    \caption{A magnified section of the shear stress - strain curve at the point (0.5W, 0.75H), with microhard boundary conditions and $L=0.2H,~l=0$, and the curve of approximate yield function $\overline{\Phi}$, showing the estimated value of first yield of 0.055}
  \label{Global-yield-diss}
  \end{figure} 
The upper bound predicts first yield to take place at a value of applied strain equal to 0.055, which may be compared with the actual value of such strain, viz. 0.056. The nature of the upper bound is clear in that $\overline{\Phi}$ at first yield has a value of 1.1. The estimated value of strain at first yield is a good approximation, albeit without a theoretical basis for estimating the sharpness of the bound. 

\section{Concluding remarks}\label{Conclusion}
In this work we have explored features of a model of strain-gradient plasticity, in the context of the model problem of a composite block in shear. The model problem, by virtue of its finite dimensions and non-homogenous composition, exhibits responses that vary with position in two directions. Microhard and microfree boundary conditions have been used; in general the responses corresponding to the two types of microscopic boundary conditions are quite distinct, the former reflecting the effects of trapping of geometrically necessary dislocations at the boundaries.  

We have given an indication of the variation in stresses with position and type of boundary condition. Similar behaviour has been observed for the two types of boundary conditions: the shear stress is dominant, and the direct stress in the direction transverse to shear has greatest magnitudes at the sides of each domains. Furthermore, the stress magnitudes are significantly higher for the microhard BC results as compared to their corresponding magnitudes in the microfree analysis. 

The elastic gap phenomenon has been illustrated for the homogeneous and domains. The gap is clearly evident, though its slope is lower than that corresponding to truly elastic behaviour, possibly as a result of the use of a viscoplastic regularization in the computations. 
 
Strengthening behaviour, corresponding to an increase in the initial yield stress, is clearly evident with increase in the dissipative length scale. This is followed by softening in the case of microfree boundary conditions, whilst for microhard BCs hardening behaviour persists through the rest of the analysis. 

Lastly, we have investigated an approximations to the global yield condition that is characteristic of the purely dissipative problem. The yield condition is given as the least upper bound of a functional involving the dissipation function, and the approximation adopted is one in which the arbitrary plastic strain is chosen to be collinear with the vector of nodal stresses. The approximation was found to give a prediction of first yield close to that observed numerically.

The problem studied in this work has provided some novel perspectives on the a model of strain-gradient plasticity. It would be useful to study this and other more complex problems further, to elucidate features that are possibly not present in one-dimensional problems. Likewise, further investigation of the yield condition would shed light on the somewhat counterintuitive notion of a global condition for yielding, suitable approximations of this global function, and its relationship to yielding as observed in numerical experiments.

\section*{Acknowledgement}\label{Acknowledgements}
The work reported in this paper was carried out with support from the National Research Foundation, through the South African Research Chair in Computational Mechanics. This support is gratefully acknowledged.

\end{document}